\documentclass[sigplan,screen]{acmart}

\AtBeginDocument{%
  }

\usepackage{array} % ?
\usepackage{enumitem}
\usepackage{subcaption}
\usepackage{algorithmic}

\usepackage{amssymb}
\usepackage{tikz, pgfplots} % ?
\usetikzlibrary{patterns}
\usetikzlibrary{positioning}
\usetikzlibrary{matrix}
\usepackage{xcolor,colortbl}
\usepackage{multirow}
\usepackage[linesnumbered,ruled,vlined]{algorithm2e}
\usepackage{makecell}
\usepackage{listings}
\usepackage{pgfplotstable} % ?
\pgfplotsset{compat=1.8}

\newcommand{\system}{Lobster\xspace}

\newcommand{\sep}[0]{~|~}
\newcommand{\ram}[0]{RAM\xspace}
\newcommand{\apm}[0]{APM\xspace}

\newcommand{\souffle}{Souffl\'{e}\xspace}

\makeatletter
\newcommand{\oeq}{\mathbin{\mathpalette\make@circled{=}}}
\newcommand{\oneq}{\mathbin{\mathpalette\make@circled{\neq}}}
\newcommand{\make@circled}[2]{%
  \ooalign{$\m@th#1\smallbigcirc{#1}$\cr\hidewidth$\m@th#1#2$\hidewidth\cr}%
}
\newcommand{\smallbigcirc}[1]{%
  \vcenter{\hbox{\scalebox{0.73}{$\m@th#1\bigcirc$}}}%
}
\makeatother

\definecolor{sclgreen}{rgb}{0,0.46,0}
\definecolor{sclblue}{rgb}{0.02,0.42,0.74}
\definecolor{scllightgrey}{rgb}{0.94,0.94,0.94}%
\definecolor{sclgreyblue}{rgb}{0.3,0.4,0.6}%
\definecolor{sclcyan}{rgb}{0.1,0.4,0.6}%
\definecolor{sclpurple}{rgb}{0.71,0,0.85}%102, 219, 255)
\definecolor{sclyellow}{rgb}{0.9,0.6,0.05}%255, 144, 33
\definecolor{sclorange}{rgb}{1,0.36,0.03}%255, 144, 33
\definecolor{sclred}{rgb}{0.6,0.2,0.0}%

\SetCommentSty{mycommfont}

\lstdefinelanguage{scallop}{
    keywords={import,type,const,rel,query,usize,where,as,String,i8,i32,i64,usize,u8,u16,u32,u64},keywordstyle=\color{blue},%
    morekeywords=[2]{and,or,not,implies,==,+,-,*,/},keywordstyle=[2]\color{sclpurple},%
    morekeywords=[3]{count,sum,prod,min,max,exists,forall,unique,top,categorical,uniform},keywordstyle=[3]\color{sclorange},%
    morecomment=[s]{/*}{*/},%
    commentstyle=\color{sclgreen},%
    morecomment=[l]{//},%
    morestring=[b]",stringstyle=\color{sclyellow}
}

\lstdefinelanguage{mypython}{
    keywords={class,def,str,return,if,elif,else,for,in,while,int,List,Tuple},keywordstyle=\color{blue},%
    morekeywords=[2]{self},
    keywordstyle=[2]\color{sclred},
    morekeywords=[3]{__init__},
    keywordstyle=[3]\color{sclcyan},
    morecomment=[s]{"""}{"""},commentstyle=\color{sclgreen},%
    morecomment=[l]{\#},%
    morestring=[b]",stringstyle=\color{sclorange}
}

\lstdefinelanguage{apm}
{
    keywords = {static, join, alloc, eval, gather, gatherif, gathern, build, overhead,loadcount,append,copy,repeat,mult,clamp,count,scan, sort, unique, merge, join, size, last},keywordstyle=\color{blue},
    commentstyle=\color{sclgreen},%
    morecomment=[l]{//},%
}

\definecolor{codegreen}{rgb}{0,0.6,0}
\definecolor{codegray}{rgb}{0.5,0.5,0.5}
\definecolor{codepurple}{rgb}{0.58,0,0.82}
\definecolor{backcolour}{rgb}{0.95,0.95,0.92}

\lstdefinestyle{mystyle}{
    commentstyle=\color{codegreen},
    keywordstyle=\color{magenta},
    numberstyle=\tiny\color{codegray},
    stringstyle=\color{codepurple},
    basicstyle=\ttfamily\footnotesize,
    breakatwhitespace=false,         
    breaklines=true,                 
    captionpos=b,                    
    keepspaces=true,                 
    showspaces=false,                
    showstringspaces=false,
    showtabs=false,                  
    tabsize=2
}

\SetKwInput{KwInput}{Input}
\SetKwInput{KwOutput}{Output}

\hypersetup{citecolor=blue,linkcolor=blue}

\definecolor{mygreen}{HTML}{D5E8D4}
\definecolor{myred}{HTML}{F8CECC}
\definecolor{myorange}{HTML}{ffcf99}
\definecolor{myblue}{HTML}{99c8f2}
\definecolor{mypurple}{HTML}{E1D5E7}
\definecolor{myyellow}{HTML}{FFF2CC}

\definecolor{mydeepgreen}{HTML}{82B366}
\definecolor{mydeepred}{HTML}{B85450}
\definecolor{mydeeporange}{HTML}{e38820}
\definecolor{mydeepblue}{HTML}{408bcf}
\definecolor{mydeeppurple}{HTML}{9673A6}
\definecolor{mydeepyellow}{HTML}{D6B656}

\definecolor{mylightblue}{HTML}{cfe2fa}
\definecolor{mylightorange}{HTML}{faeccf}

\copyrightyear{2026}
\acmYear{2026}
\setcopyright{cc}
\setcctype{by}
\acmConference[ASPLOS '26]{Proceedings of the 31st ACM International Conference on Architectural Support for Programming Languages and Operating Systems, Volume 1}{March 22--26, 2026}{Pittsburgh, PA, USA}
\acmBooktitle{Proceedings of the 31st ACM International Conference on Architectural Support for Programming Languages and Operating Systems, Volume 1 (ASPLOS '26), March 22--26, 2026, Pittsburgh, PA, USA}
\acmPrice{}
\acmDOI{10.1145/3760250.3762232}
\acmISBN{979-8-4007-2165-6/26/03}

\begin{document}

%% The "title" command has an optional parameter,
%% allowing the author to define a "short title" to be used in page headers.
\title{Lobster: A GPU-Accelerated Framework for Neurosymbolic Programming}

\author{Paul Biberstein}
\orcid{1234-5678-9012}
\email{paulbib@seas.upenn.edu}
\affiliation{%
  \institution{University of Pennsylvania}
  \city{Philadelphia}
  \state{Pennsylvania}
  \country{USA}
}
\author{Ziyang Li}
\orcid{0000-0003-3925-9549}
\email{liby99@seas.upenn.edu}
\affiliation{%
  \institution{Johns Hopkins University}
  \city{Baltimore}
  \state{Maryland}
  \country{USA}
}
\author{Joseph Devietti}
\orcid{0000-0002-9330-7233}
\email{devietti@seas.upenn.edu}
\affiliation{%
  \institution{University of Pennsylvania}
  \city{Philadelphia}
  \state{Pennsylvania}
  \country{USA}
}
\author{Mayur Naik}
\orcid{0000-0003-1348-8618}
\email{mhnaik@seas.upenn.edu}
\affiliation{%
  \institution{University of Pennsylvania}
  \city{Philadelphia}
  \state{Pennsylvania}
  \country{USA}
}

\renewcommand{\shortauthors}{Paul Biberstein, Ziyang Li, Joseph Devietti, and Mayur Naik}

\begin{abstract}

Neurosymbolic programs combine deep learning with symbolic reasoning to achieve better data efficiency, interpretability, and generalizability compared to standalone deep learning approaches.  
However, existing neurosymbolic learning frameworks implement an uneasy marriage between a highly scalable, GPU-accelerated neural component and a slower symbolic component that runs on CPUs.

We propose Lobster, a unified framework for harnessing GPUs in an end-to-end manner for neurosymbolic learning.
Lobster maps a general neurosymbolic language based on Datalog to the GPU programming paradigm.
This mapping is implemented via compilation to a new intermediate language called APM.
The extra abstraction provided by apm allows Lobster to be both flexible, supporting discrete, probabilistic, and differentiable modes of reasoning on GPU hardware with a library of provenance semirings, and performant, implementing new optimization passes.

We demonstrate that Lobster programs can solve interesting problems spanning the domains of natural language processing, image processing, program reasoning, bioinformatics, and planning.
On a suite of 9 applications, 
Lobster achieves an average speedup of 3.9x over Scallop, a state-of-the-art neurosymbolic framework, and enables scaling of neurosymbolic solutions to previously infeasible tasks.
\end{abstract}




\maketitle

\section{Introduction}
\label{sec:introduction}

Deep learning and classical algorithms represent two predominant paradigms of modern programming.
Classical algorithms excel at problems with clearly defined rules and structured data, such as sorting a list of numbers or finding a shortest path in a graph. 
In contrast, deep learning is well suited to contexts where classical algorithmic approaches become intractable, particularly for problems involving noisy, complex, and high-dimensional data---such as detecting objects in an image or parsing natural language text.

Many machine learning problems in different domains demand the complementary capabilities of these two paradigms.
Neurosymbolic programming \cite{chaudhuri2021neurosymbolic} is an emerging approach to solve such problems by
suitably decomposing the computation between a neural network and a symbolic program.
The resulting \textit{neurosymbolic programs} have been demonstrated to achieve better data efficiency, interpretability, and generalizability compared to standalone deep learning approaches.
Such properties are crucial for various safety-critical domains such as system security~\cite{li2025iris}, cyber-physical systems~\cite{zheng2025neurostrata}, and healthcare~\cite{wu2024discret}.

\begin{figure}[t]
    \centering
    \includegraphics[width=0.48\textwidth]{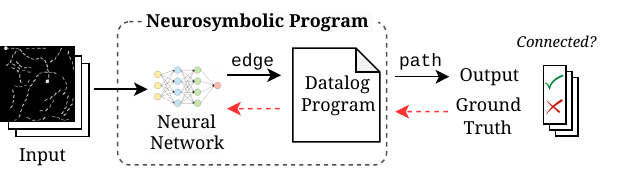}
    \caption{
        An example neurosymbolic program.
    }
    \label{fig:overview-neurosym}
    \vspace{-0.2in}
\end{figure}

Recent frameworks such as DeepProbLog \cite{manhaeve2018deepproblog}, Scallop \cite{li2023scallop}, and ISED \cite{solko2024data} have enhanced the programmability and accessibility of neurosymbolic applications. 
\autoref{fig:overview-neurosym} illustrates a neurosymbolic program for solving a binary image-classification problem \cite{tay2020longrange}.
The symbolic program is specified in Datalog \cite{abiteboul1995foundations}, a declarative language. 
Crucially, by using a differentiable Datalog engine, gradients can be back-propagated through the program to train the neural network, thereby enabling automatic learning of relevant image features without manual engineering.

\begin{figure*}
        \centering
        \includegraphics[width=1.0\textwidth]{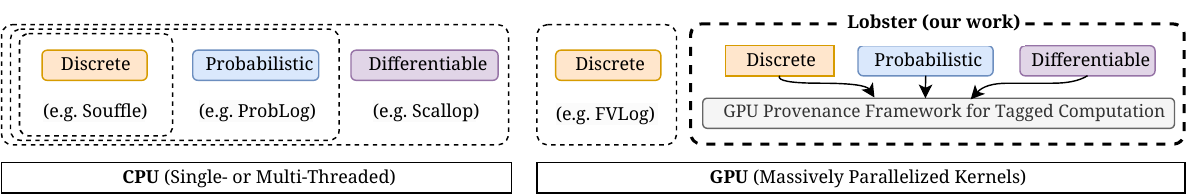}
    \caption{High-level comparison between \system and existing frameworks.}
    \label{fig:overview}
\end{figure*}

Despite their benefits, neurosymbolic programs using these frameworks incur significant computational overhead during training and inference.
The scalability challenges stem primarily from managing probabilistic data and maintaining end-to-end differentiability. 
For instance, in the neurosymbolic program in \autoref{fig:overview-neurosym}, a neural network identifies edges in an image. These edges are represented in an {\tt edge} relation, where each tuple has an associated probability representing the network's confidence. The symbolic program computes the transitive closure to produce a {\tt path} relation, where the probability of each path tuple must take into account all the possible ways to derive it from the edge tuples, and their associated probabilities.
Since it is often intractable to perform exact probabilistic reasoning, approximated probabilistic inference is employed, though this has only limited scalability benefits.
Differentiability further complicates the problem by requiring us to track each input’s contribution to the output, increasing space and time complexity due to the extra book-keeping required for gradients.

In this paper, we propose \system, a GPU-accelerated framework designed to enhance the scalability of neurosymbolic programming.
\system's core innovation is efficiently mapping Datalog---a logic programming language shown to be effective in neurosymbolic contexts \cite{li2023scallop}---onto GPU architectures, for different modes of reasoning: discrete, probabilistic, and differentiable.
The key architectural choice that enables the efficiency of this mapping is compilation to a new intermediate language, called \apm, which is expressive enough to support complex reasoning but at the same time restricted enough to ensure massively parallel execution.

Supporting both advanced reasoning modes and GPU acceleration makes \system the first system of its kind, as shown in \autoref{fig:overview}.
While various engines exist for discrete \cite{bernhard2016souffle}, probabilistic \cite{anton2015problog2}, and differentiable \cite{li2023scallop} settings, they are limited to CPU runtimes with single- or multi-threading. 
Other work \cite{shovon2023gpujoin,sun2024moderndataloggpu} implements GPU-accelerated Datalog execution, but does not support differentiable and probabilistic reasoning.
In contrast, \system handles general neurosymbolic queries with multiple reasoning modes within a unified, GPU-accelerated framework of {\em provenance semirings}~\cite{provenancesemiring}.
This requires fundamental changes to the program semantics compared to discrete Datalog, enriching the runtime with semiring-based tags that propagate alongside data throughout the computation.
To support this efficiently on modern hardware, \system introduces GPU-optimized operators specifically tailored for tagged computation.
By supporting a library of 7 common semirings in the literature, \system allows employing reasoning in a particular mode (e.g. probabilistic) by simply selecting a suitable corresponding semiring (e.g. Top-$k$-Proofs).

In addition to describing the compilation to and execution of \apm, we propose a number of optimizations unique to \system and discuss considerations for our implementation---a fully-fledged compiler and runtime written in Rust---that can execute existing Datalog-based neurosymbolic programs on GPUs without any modifications.

In summary, the core contributions of this paper are:
\begin{itemize}[topsep=0pt, left=5pt]
    \item We introduce \system, the first GPU-accelerated neurosymbolic programming framework.
    \item We propose the \apm language and show how to compile Datalog to \apm.
    \item We implement a compiler and runtime for Lobster using Rust and CUDA.
    \item We evaluate \system on an extensive set of discrete, probabilistic, and differentiable benchmarks, showing that \system consistently outperforms prior systems, including more specialized ones. \system achieves a speedup of 3.9x on average and upto >100$\times$ over Scallop~\cite{scallop}, the closest existing state-of-the-art system.
\end{itemize}

The code for \system is publicly available at \url{https://github.com/P-bibs/Lobster}.

The rest of the paper is organized as follows. 
We first give an illustrative overview of \system in Section~\ref{sec:motivating}. 
Then we introduce \system's core language and compiler (Section~\ref{sec:language-compiler}), optimizations (Section~\ref{sec:optimizations}), and implementation details (Section~\ref{sec:implementation}).
Finally, we present experimental results (Section~\ref{sec:evaluation}) and related work (\autoref{sec:related}).

\section{Illustrative Overview}
\label{sec:motivating}

\begin{figure*}
    \begin{subfigure}[c]{0.45\textwidth}
        \centering
        \resizebox{1.0\textwidth}{!}{
            \begin{tikzpicture}
                \node[anchor=south west, inner sep=0] (img1) at (0,0) {\includegraphics[width=3cm]{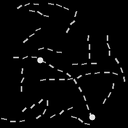}};
                \draw[->, thick] (3.05,1.5) -- (3.45,1.5);
                
                \node[anchor=south west, inner sep=0] (img2) at (3.5,0) {\includegraphics[width=3cm]{medium_connected.png}};
                \begin{scope}
                    \clip (img2.south west) rectangle (img2.north east);
                    \draw[step=0.333cm,gray,very thin,opacity=0.6,shift={(0.166,0)}] (3, 0) grid (6.5, 3);
                \end{scope}
                
                \draw[->, thick] (6.55,1.5) -- (6.95,1.5);

                \node[anchor=south west, inner sep=0] (img3) at (7,0) {\includegraphics[width=3cm]{medium_connected.png}};
                \begin{scope}
                    \clip (img3.south west) rectangle (img3.north east);
                    \def\offset{0.1666}
                    \foreach \x [count=\xi] in {7,7.333,...,11} { 
                        \foreach \y [count=\yi] in {0,0.333,...,3} {
                            \def\nodecolor{red}
                            \ifthenelse{\xi=3 \and \yi=5}{\def\nodecolor{green}}{}
                            \ifthenelse{\xi=7 \and \yi=1}{\def\nodecolor{green}}{}
                            \node[mark size=1pt, color=\nodecolor] 
                                (node\xi\yi) at (\x+\offset, \y+\offset) {\pgfuseplotmark{*}};
                        }
                    }
                    \foreach \from/\to in {35/45, 45/55, 55/54, 54/64, 64/63, 63/62, 62/72, 72/71} {
                        \draw[->, white] (node\from.center) -- (node\to.center);
                    }
                \end{scope}
            \end{tikzpicture}
        }
        \caption{
            An example graph a trained model extracts for the Pathfinder task. Green vertices are endpoints and white edges are predicted connectivity. Additional edges not relevant to endpoint connectivity are elided for clarity.
        }
        \label{fig:motivating-pipeline}
    \end{subfigure}
    \hfill
    \begin{subfigure}[c]{0.5\textwidth}
        \centering
        \includegraphics[width=1.0\textwidth]{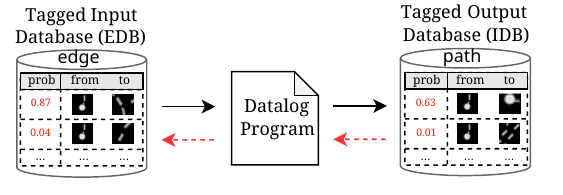}
        \caption{
            The input and output for the Pathfinder Datalog program.
        }
        \label{fig:overview-database}
    \end{subfigure}
    \begin{subfigure}{\textwidth}
        \begin{subfigure}[c]{0.45\textwidth}
            \begin{lstlisting}[
                language=scallop,
                basicstyle=\footnotesize\tt,
                xleftmargin=0.6cm,
                numbers=left,
            ]
type Cell = u32
type edge(x: Cell, y: Cell)
type is_endpoint(x: Cell)

rel path(x, y) :-
  edge(x,y) or (path(x, z) and edge(z, y)).
rel endpoints_connected() :- is_endpoint(x),
  is_endpoint(y), path(x, y), x != y.
            \end{lstlisting}
            \caption{
                A \system program that computes reachability information for a graph extracted by a neural network.
            }
            \label{fig:motivating-program}
        \end{subfigure}
        \hfill
        \begin{subfigure}[c]{0.47\textwidth}
            \begin{subfigure}{\textwidth}
                \centering
                \small
                \begin{tabular}{l|ll}
                Accuracy       & Neural   & Neurosymbolic    \\ \hline
                Pathfinder     & 71.40   &  87.42      \\
                Pathfinder-x   & 49.80   &  89.46
                \end{tabular}
                \caption{Accuracy of neural methods vs. neurosymbolic methods~\cite{scallop}. ``Pathfinder-x'' is a more challenging dataset.}
                \label{fig:motivating-accuracy}
            \end{subfigure}
            \begin{subfigure}{\textwidth}
                \vspace{0.2in}
                \centering
                \small
                \begin{tabular}{l|ll}
                               & Scallop     & \system    \\ \hline
                Training Time  & 41 hr.      & 32 hr      \\
                \end{tabular}
                \caption{Training time of Scallop versus \system.}
                \label{fig:motivating-speedup}
            \end{subfigure}
        \end{subfigure}
    \end{subfigure}
    \caption{The overall pipeline, symbolic program, and the acceleration result of using \system for the Pathfinder task.}
    \label{fig:motivating-example}
\end{figure*}

We illustrate \system using an example image-reasoning task Pathfinder~\cite{tay2020longrange} wherein the goal is to determine whether two dots in the input image are connected by dashed lines (Figure~\ref{fig:motivating-example}).
Neurosymbolic methods have been shown to achieve greater accuracy than purely neural methods on this task~\cite{scallop}, but the symbolic performance bottleneck quickly appears as the lengths and complexities of lines (and therefore reasoning chain size) increase.

\subsection{A Neurosymbolic Solution}

As with many visual reasoning tasks, the Pathfinder task presents an obvious opportunity for neural/symbolic decomposition: a neural network can extract a structured representation from the image, and a symbolic engine can be used to reason over this representation.

More formally, we choose a discretization factor $n$ and overlay the lattice graph $G_{n,n}$ on the image, where each vertex corresponds to a square region of pixels.
A convolutional neural network model~\cite{lecun1989backprop} predicts probabilities for each edge in $G_{n,n}$, with high probability indicating confidence that two grid cells are connected by a dashed line.
Additionally, the model predicts a probability corresponding to the presence of a dot in each grid cell.
The symbolic component then computes the transitive closure of the graph and queries the result to determine the probability two dots are connected.

\subsection{A Scallop Implementation}
We present the high-level pipeline of implementing a neurosmbolic solution using Scallop, a state-of-the-art neurosymbolic framework, in Figure~\ref{fig:motivating-pipeline}. 
We specify the symbolic component as a Datalog program which is shown in Figure~\ref{fig:motivating-program}.
Datalog offers an intuitive interface for defining data types (line~1) and relation types (lines~2-3) which we use to specify the inputs. 
Notice that the graph is encoded as a binary relation between \texttt{Cell}s, which we represent with unsigned integers.
Our program contains recursive rules, which are key for modeling reachability in a graph.
The actual reachability rule is defined succinctly in line~6.

Notably, the probabilistic reasoning semantics is abstracted away.
The \texttt{edge} relation in Figure~\ref{fig:motivating-program} (line~2) contains probabilistic facts extracted by the underlying neural network.
For example, a fact $0.97$\texttt{::edge(0, 1)} represents a prediction that a dotted line connects cell 0 and cell 1 with the probability of $0.97$.
All the derived facts, such as \texttt{path(i, j)}, will carry probabilities computed from a concrete set of edges that are used to derive them.
A proof for \texttt{path(1,5)}, for example, could be the conjunction $\texttt{is\_endpoint(1)} \wedge \texttt{edge(1, 2)} \wedge \dots \wedge \texttt{edge(4, 5)} \wedge \texttt{is\_endpoint(5)}$, representing a concrete path formed by a subset of the probabilistic input facts.
This computation is manifested by the underlying provenance framework within \system.
In our solution, we use a probabilistic provenance called \texttt{diff-top-1-proofs} which instruments the program to carry the most-likely path between any two cells.

Since the symbolic engine is differentiable, the entire neurosymbolic system can be trained end-to-end, resulting in a trained neural network despite only having ``yes''/``no'' supervision for each image. The neurosymbolic solution achieves 87.42\% accuracy on the Pathfinder task, surpassing the 71.40\% accuracy of purely neural methods (Figure~\ref{fig:motivating-accuracy}).

\subsection{Scalability and Programmability Challenges}
The neurosymbolic solution poses a significant scalability challenge. 
While the neural component can utilize modern hardware accelerators like GPUs and TPUs, the symbolic component runs on CPUs alone. 
This presents a performance bottleneck, as in order to calculate derivatives, the symbolic engine must consider all possible predicted graphs and their associated probabilities. 
As the size of the input image or difficulty of line curvature increases, the number of possible structures and the size of their associated weights also grows exponentially, leading to a combinatorial explosion in the number of required computations.

An expert could write custom GPU kernels to accelerate this specific task, but this requires specialized knowledge of CUDA performance tuning. Instead, we seek to build a general, GPU-accelerated framework that accelerates any logic program used as part of a neurosymbolic pipeline, in order to make GPU acceleration widely accessible.

\subsection{Our Results}

Figure~\ref{fig:motivating-speedup} shows the speedup of \system over the CPU baseline, Scallop, on the Pathfinder task.
Scallop spends significant time on symbolic computation, while \system can perform the requisite combinatorial graph processing much quicker.
Importantly, users do not need to change existing neurosymbolic programs to benefit from the acceleration.
\system's efficiency gains are enabled by mapping a significant fragment of Datalog to the GPU programming paradigm, and making judicious decisions for representing relations, parallelizing relational operators, and scheduling computation which we describe next.

\paragraph{How to Represent Relations?}
\system uses a flat, column-oriented layout in order to make optimal use of the GPU memory hierarchy.
While performant CPU Datalog engines such as \souffle make use of multi-layer data structures such as B-trees~\cite{souffle-progam-analysis}, our simple layout makes more sense in the context of GPU acceleration as column-oriented layouts are cache-friendly, and GPU programs are often memory-bound rather than compute-bound.
Concretely, this means some of the simpler operations in the Pathfinder task, such as \texttt{union}ing the relation \texttt{edge} with the relation \texttt{path} to initially populate \texttt{path}, can reach close to 100\% utilization of the GPU memory bus.
This choice also suits the context of executing Datalog programs. For instance, the transitive closure operation is compiled to a series of query operations consisting of relational operations like \emph{join} and \emph{project} which operate on specific columns.
As a result, columnar data allows more natural algorithm implementations.
We discuss the memory layout further in Section~\ref{sec:implementation}.

\paragraph{How to Parallelize Relational Operators?}
Beyond memory layout, efficient algorithms are necessary to improve the performance of symbolic computations.
In \system, the key insight is that Datalog programs are compiled to a core set of relational queries, and each of these queries can be individually parallelized to improve their performance on potentially massive inputs.
For example, the rule on line 6 of Figure~\ref{fig:motivating-program} is compiled to a query that includes joining the entire set of currently discovered paths against the base edge set.
The size of the input to this join grows exponentially as the program iterates, so executing the join with data-parallelism with respect to its input is critical.
We describe how our compiler exposes this parallelism in Section~\ref{sec:language-compiler}.

\paragraph{How to Handle Provenance Tags?}
Computing derivatives of the symbolic computation is necessary for training the neural network via gradient descent. This necessitates tracking the provenance of each fact in the Datalog program which involves computing over large and potentially complicated semiring tags. 
To adapt powerful but complex
semirings like diff-top-k-proofs~\cite{li2023scallop}---which tracks up to $k$ proofs of arbitrary size for each fact---to the GPU, we leverage two insights.
First, we observe that the max proof size can be statically determined at compile time. 
Second, we find that the \texttt{diff-top-1-proofs} semiring that tracks just one proof is sufficient for many programs; \system could also easily be extended to track larger $k$ as well.

\section{Language and Compiler}
\label{sec:language-compiler}

\newcommand{\instr}[1]{{\color{blue}\texttt{#1}}}
\newcommand*{\mline}[1]{%
\begingroup
    \renewcommand*{\arraystretch}{1.2}%
    % hard-coded 12cm width on mlines
   \begin{tabular}[c]{@{}>{\raggedright\arraybackslash}p{12cm}@{}}#1\end{tabular}%
  \endgroup
}

\begin{table*}
    \caption{A summary of instructions in the \apm language. Lowercase latin characters represent registers, while capital latin characters represent scalar integers. We write $\overline{r_n}$ to denote a sequence of $n$ registers and $r_t$ to denote a register containing semiring tags. $\alpha_{n,m}$ denotes a function from $n$-tuples to $m$-tuples, $\sigma$ a function $\tau \times \tau \to \tau$ for some type $\tau$, and $\rho$ a relation in the database.}
    \label{fig:apm-summary}
    \small
    \begin{tabular}{l|l}
        \textbf{Signature} & \multicolumn{1}{c}{\bfseries Description} \\
        \hline
        $\instr{alloc}\langle \tau_1,\ldots,\tau_n \rangle(\overline{r_n}, S)$                               & \mline{Allocate $n$ registers each of size $S$ and types $\tau_1,\ldots,\tau_n$. In practice, the type can be inferred based on usage and is elided.} \\
        $\overline{d_m} \gets \instr{eval}\langle \alpha_{n,m} \rangle (\overline{s_n})$                 & \mline{Evaluate $\alpha$ on each row of $\overline{s_n}$.} \\
        $\overline{d_n} \gets \instr{gather}(i, \overline{s_n})$                         & \mline{Gather rows of $\overline{s_n}$ based on indices $i$.} \\
        $d \gets \instr{gather}\langle \alpha_{n,1} \rangle(\overline{i_n}, \overline{s_n})$  & \mline{Gather rows of $\overline{s_n}$ based on indices $\overline{i_n}$ and reduce the resulting tuple with $\alpha$.} \\
        $\instr{store} \langle \rho\rangle(\overline{s_n}, s_t)$                                         & \mline{Store registers $\overline{s_n}$ and $s_t$ as the columns and tags for relation $\rho$ with arity $n$ in the database.} \\
        $\left[\overline{s_n}, s_t\right] = \instr{load}\langle \rho \rangle()$                                     & \mline{Loads the columns and tags of relation $\rho$ with arity $n$ from the database into registers $\overline{s_n}$ and $s_t$.} \\
        $ d \gets \instr{build} (\overline{s_n})$                                             & \mline{Builds a hash index for the table with columns $\overline{s_n}$.} \\
        $\overline{d_n} \gets \instr{count} (\overline{b_n}, h, \overline{a_n})$                        & \mline{Count the number of occurrences of each tuple in the table with columns $\overline{b_n}$ in the table with columns $\overline{a_n}$ via the hash index $h$.} \\
        $\overline{d_n} \gets \instr{scan}(s)$                                                             & \mline{Computes the exclusive prefix sum of register $s$.} \\
        $\left[d_l, d_r\right] \gets \instr{join} \langle W \rangle(\overline{b_m}, \overline{a_n}, h, c, o)$            & \mline{Produces the resulting indices from a $W$ column join of two tables with columns $\overline{b_m}$ and $\overline{a_n}$ via the hash index $h$, histogram $c$, and histogram prefix sum $o$.} \\
        $\overline{d_n} \gets \instr{copy} (\overline{s_n}) $                                            & \mline{Copies from register $\overline{s_n}$, truncating if the destination is smaller than the source.} \\
        $\overline{d_n} \gets \instr{sort} (\overline{s_n}) $                                            & \mline{Lexicographically sorts the table with columns $\overline{s_n}$.} \\
        $\left[\overline{d_n}, s\right] \gets \instr{unique}\langle \sigma \rangle(\overline{s_n})$                                       & \mline{Merges adjacent duplicate rows via $\sigma$ from the table with columns $\overline{s_n}$, returning the number of unique elements $s$.} \\
        $\overline{d_n} \gets \instr{merge}(\overline{a_n}, \overline{b_n})$                             & \mline{Merges two lexicographically sorted tables with columns $\overline{a_n}$ and $\overline{b_n}$.} \\
    \end{tabular}
\end{table*}

\system focuses on accelerating the Datalog back-end with GPU hardware.
This poses a challenge, as the process of supporting rich reasoning is at odds with ensuring high performance.
To achieve both of these goals, Lobster introduces a new intermediate language, \apm (Abstract Parallel Machine), which is designed to simplify the process of compiling and optimizing Datalog programs for GPU execution.
We assume an existing Datalog compiler is capable of converting a user-level program to a mid-level program based on relational algebra. 
From there, \system compiles the relational algebra down to an APM program that can be executed on the GPU.
In this section, we describe the low-level sequential language \apm and present the compilation process from the mid-level relational algebra language to \apm.

\subsection{Background}

\paragraph{Relational Algebra Machine}
\label{subsec:ram}
We start by describing our compiler's source language, the Relational Algebra Machine (RAM), which is based on the familiar language of \textit{Relational Algebra} for expressing database queries~\cite{abiteboul1995foundations}.
The abstract syntax of RAM is shown in Figure~\ref{fig:ram-syntax}.
At a high level, executing a \ram program $\bar{\phi}$ means sequentially executing each stratum $\phi_1, \dots, \phi_n$.
Within each stratum, rules are iteratively applied to the extensional database (EDB), which contains the input facts, until a fix-point is reached.
The newly derived facts form the intensional database (IDB), which accumulates intermediate results produced by the program.
Each rule $\rho \leftarrow \epsilon$ consists of a target relation $\rho$ and a query $\epsilon$. This query is a dataflow graph with  many sources but only one sink. 
The operators in the graph are a core fragment of relational algebra operators, comprising project ($\pi$), select ($\sigma$), and join ($\bowtie$) as well as three set operators, union ($\cup$), product ($\times$), and intersect ($\cap$). 
Note that $\pi$ and $\sigma$ allow taking arbitrary projection or selection functions, while the join operation $\bowtie$ accepts the number of columns to perform join on.
For this section, we focus on accelerating a single recursive stratum.

\begin{figure}
    \begin{minipage}[b]{0.42\textwidth}
        \centering
        \footnotesize
        \[
            \begin{array}{rcrl}
            \text{(Predicate)} & \rho \\
            \text{(Projection Fn.)} & \alpha \\
            \text{(Selection Fn.)} & \beta \\
            \text{(Expression)} & \epsilon & ::=  & \rho \ \ \sep\ \ \pi_\alpha(\epsilon)\ \ \sep\ \ \sigma_\beta(\epsilon)\ \ \sep\ \ \epsilon_1 \bowtie_n \epsilon_2 \\
                                &   & \sep & \epsilon_1 \cup \epsilon_2\ \ \sep\ \ \epsilon_1 \times \epsilon_2\ \ \sep\ \ \epsilon_1 \cap \epsilon_2   \\
            \text{(Rule)} & \psi & ::= & \rho \leftarrow \epsilon \\
            \text{(Stratum)} & \phi & ::= & \{\ \psi_1, \dots, \psi_n\ \} \\
            \text{(Program)} & \overline{\phi} & ::= & \phi_1; \dots; \phi_n \\
            \end{array}
        \]
        \captionof{figure}{The \ram language.}
        \label{fig:ram-syntax}
    \end{minipage}
\end{figure}

\paragraph{Provenance Semirings}
Relational algebra programs can incorporate differentiable or probabilistic reasoning by tagging each fact with additional information such as probabilities or boolean formulas, as shown in prior work~\cite{manhaeve2018deepproblog,li2023scallop}.
More generally, \textit{provenance semirings} \cite{provenancesemiring} enable programmable semantics that allow tags from an arbitrary semiring.
Formally speaking, a provenance semiring $T$ is a 5-tuple $(T, \textbf{0}, \textbf{1}, \oplus, \otimes)$ where $T$ is the space of tags (Figure~\ref{fig:prov-semiring}). 
$\oplus$ and $\otimes$ dictate how tags are combined through disjunction and conjunction operations.
In Figure~\ref{fig:prov-example}, we show a few provenance semirings used in the literature \cite{provenancesemiring,anton2015problog2,huang2021scallop} for discrete reasoning and approximated probabilistic reasoning.
As an example, a tag can be a boolean formula $\phi \in \Phi$ represented in disjunctive normal form (DNF) under set notation.
Here, the boolean variables $\nu$ will be references to facts in the input databases.
With probability $\Pr(\nu)$ attached, one might perform top-$k$ filtering on proofs to avoid blow-up of the boolean formulas.
In order to support the discrete, probabilistic, and differentiable modes of reasoning, \system implements 7 commonly used provenance semirings, which we elaborate in Section~\ref{sec:provenance-semiring}.

\begin{figure}
    \footnotesize
    \begin{subfigure}{\linewidth}
        \centering
        \[
            \begin{array}{rrcl}
              \text{(Tag)} & t & \in & T \\
              \text{(False)} & \mathbf{0} & \in & T \\
              \text{(True)} & \mathbf{1} & \in & T \\
              \text{(Disjunction)} & \oplus & : & T \times T \rightarrow T \\
              \text{(Conjunction)} & \otimes & : & T \times T \rightarrow T
            \end{array}
        \]
        \caption{
            The provenance semiring structure.
        }
        \label{fig:prov-semiring}
    \end{subfigure}
    \begin{subfigure}{\linewidth}
        \centering
        \vspace{0.1in}
        \begin{tabular}{cccccc}
            \toprule
            Provenance & $T$ & $\mathbf{0}$ & $\mathbf{1}$ & $\oplus$ & $\otimes$ \\
            \midrule
            Bool & $\{\bot, \top\}$ & $\bot$ & $\top$ & $\vee$ & $\wedge$
            \\
            Max-Min-Prob & $[0, 1]$ & $0$ & $1$ & $\text{max}$ & $\text{min}$
            \\
            Top-$k$-Proofs & $\Phi$ & $\{\}$ & $\{\emptyset\}$ & $\vee_k$ & $\wedge_k$
            \\
            \bottomrule
        \end{tabular}
        \caption{
            Common examples of provenance semirings.
        }
        \label{fig:prov-example}
    \end{subfigure}
    \caption{Provenance semiring structure and examples.}
    \vspace{-0.2in}
\end{figure}

\begin{figure*}[t]

    \begin{subfigure}{\textwidth}
        \centering
        \scriptsize
        \pgfdeclarelayer{background layer}
        \pgfsetlayers{background layer,main}
        \begin{tikzpicture}[
            level distance=0.7cm,
            level 1/.style={sibling distance=2.0cm},
            level 2/.style={sibling distance=2.0cm},
            every node/.style={draw, fill=white},
            codebox/.style={inner sep=5pt},
            pill/.style={rounded corners=2mm},
        ]
            \node (t5) {$\pi_{\lambda (i,j,k).(k,j)}$}
                child {node (t4) {$\bowtie_1$}
                    child {node (t2) {$\pi_{\lambda (i,j).(j,i)}$}
                        child {node[pill] (t1) {$\texttt{path}(x,z)$}
                        }
                    }
                    child {node[pill] (t3) {$\texttt{edge}(z,y)$}
                    }
                };

            \node[left=0.0cm of t1,draw=none] {\bf\texttt{r1}};
            \node[left=0.0cm of t2,draw=none] {\bf\texttt{r2}};
            \node[left=0.0cm of t3,draw=none] {\bf\texttt{r3}};
            \node[left=0.0cm of t4,draw=none] {\bf\texttt{r4}};

            \node[codebox, minimum width=6.5cm, above=0.4cm of t5, text width=6.5cm, ] {
\begin{lstlisting}[aboveskip=-4px,escapechar=\%,language=scallop,basicstyle=\scriptsize\tt]
rel path(x,y) = path(x,z) and edge(z,y)
\end{lstlisting}
};

            \node[codebox, minimum width=6.2cm, right=2.8cm of t3, text width=6.2cm, yshift=1cm] (code1) {%
\begin{lstlisting}[aboveskip=-4px,escapechar=\%,mathescape=true,language=apm,basicstyle=\scriptsize\tt]
alloc($\textbf{h}$,$\text{size}(\textbf{r3}_1)*\mathcal{O}$)
static $h$ $\gets$ build($[\textbf{r3}_1]$)
alloc([$\textbf{c},\textbf{o}$],size($\textbf{r2}_1$))
$c$ $\gets$ count($[\textbf{r2}_1]$,$\textbf{h}$,$[\textbf{r3}_1]$)
$o$ $\gets$ scan($\textbf{c}$)
alloc([$\textbf{i}_l$,$\textbf{i}_r$,$\textbf{r4}_1,\textbf{r4}_2,\textbf{r4}_3,\textbf{r4}_t$],last($\textbf{o}$))
$[\textbf{i}_l,\textbf{i}_r]$ $\gets$ join$\langle 1\rangle$($[\textbf{r2}_1,\textbf{r2}_2]$,$[\textbf{r3}_1,\textbf{r3}_2]$,$\textbf{h}$,$\textbf{c}$,$\textbf{o}$)
$[\textbf{r4}_1,\textbf{r4}_2]$ $\gets$ gather($\textbf{i}_l$,$[\textbf{r3}_1,\textbf{r3}_2]$)
$[\textbf{r4}_3]$ $\gets$ gather($\textbf{i}_r$,$[\textbf{r2}_2]$)
$\textbf{r4}_t$ $\gets$ gather$\langle \otimes \rangle$([$\textbf{i}_l$,$\textbf{i}_r$],$[\textbf{r3}_t,\textbf{r2}_t]$)
\end{lstlisting}};

            \node[codebox, minimum width=6.2cm, below=0.2cm of code1, text width=6.2cm, ] (code2) {%
\begin{lstlisting}[aboveskip=-4px,escapechar=\%,mathescape=true,language=apm,basicstyle=\scriptsize\tt]
alloc($[\textbf{r2}_1,\textbf{r2}_2,\textbf{r2}_t]$, size($\textbf{r1}_1$))
$[\textbf{r2}_1,\textbf{r2}_2]$ $\gets$ eval$\langle\lambda(i,j),(j,i)\rangle$ ($[\textbf{r1}_1,\textbf{r1}_2]$)
$[\textbf{r2}_t]$ $\gets$ copy($\textbf{r1}_t$)
\end{lstlisting}};

            \begin{pgfonlayer}{background layer}
                \draw[-, dashed] (t4.north east) -- (code1.north west);
                \draw[-, dashed] (t4.south east) -- (code1.south west);
                \draw[-, dashed] (t2.north east) -- (code2.north west);
                \draw[-, dashed] (t2.south east) -- (code2.south west);
            \end{pgfonlayer}

        \end{tikzpicture}
    \end{subfigure}

    \caption{
        In this example, we compile a part of the rule shown in Figure~\ref{fig:motivating-program} (line~6).
        The code block on the top shows the Datalog rule, while bottom-left illustrates the abstract syntax tree of the \ram program compiled from it.
        We expand the nodes \textbf{\texttt{r2}} and \textbf{\texttt{r4}} on the right to show their low-level \apm code.  We denote with $\mathcal{O}$ a configurable parameter that determines the size of the hash table used in the join implementation.
    }
    \label{fig:compilation-example}
\end{figure*}

\subsection{\apm: A Language for Parallel Machines}
\label{subsec:apm}
\apm is a low-level, assembly-style procedural language that explicitly exposes allocations and is composed exclusively of instructions which permit massively parallel execution.
An overview of the instructions is provided in Table~\ref{fig:apm-summary}.
\apm seeks to solve the problem that, traditionally, GPU programming is much like C programming: an unbounded set of programs can be expressed, even ones that map poorly to the underlying hardware.
\apm alleviates this problem by taking the implicit guidelines of the GPU programming model and making them explicit in the design of \apm.
This results in a desirable property: once a program is compiled to \apm, efficient GPU execution is assured.
We consider a number of core limitations of the GPU programming paradigm and make a corresponding design decision in \apm:
\begin{enumerate}[leftmargin=*]
    \item \textbf{Lockstep Execution}
    While GPUs have thousands of cores available for parallel computation, these cores are not as flexible as CPU cores.
Specifically, GPU cores implement a single instruction, multiple data (SIMD) paradigm, in which a set of 32 threads (known as a \textit{warp}) must execute the same set of instructions while operating over separate thread registers.
This informs the \textbf{lack of control flow} in \apm, ensuring minimal thread divergence.

    \item \textbf{Allocation}
    Allocating GPU memory while GPU code is executing has negative performance implications.
    Therefore, data structures commonly used in database systems that rely on pointer chasing like B-Trees and Tries are non-starters in programs wishing to execute on GPUs.
    Instead, data structures like sorted arrays, which use large contiguous blocks of memory and can pre-allocate enough memory for their use up-front, are preferred.
    This restriction is respected by requiring \apm programs be in \textbf{static single assignment} (SSA) form~\cite{cytron1991efficiently}, and by requiring \textbf{all registers be explicitly allocated} with a size before use.
    As a result, compiling to \apm requires statically determining memory allocation points, and determining register lifetimes is trivial.

    \item \textbf{Coalesced Memory}
    In GPUs, memory accesses are fastest when threads within a warp access consecutive memory locations, a pattern known as coalesced memory access.
    As such, a columnar representation for relational tables helps ensure maximum utilization of GPU memory bandwidth by ensuring the common path of per-column memory operations results in coalesced accesses.
    Correspondingly, all registers in \apm are \textbf{vector registers} that store a non-resizable buffer of identically-typed values. 

\end{enumerate}

\subsection{Compiling \ram to \apm}
\label{subsec{ram-to-apm}}

Given the design considerations of \apm, we now must determine how to compile a \ram program to \apm.
The most important decision is determining how to represent tables in \apm, as they are the base unit of data in relational algebra.
As tables consist of a fixed schema of columns of identical size, it is straightforward to represent an arity $n$ relation as $n$ registers of equal size, adding an additional register for provenance semiring tags.
It is then sensible to discuss sets of registers in \apm as tables, just as we discuss sets of facts in \ram as relations.

With table representation fixed, compiling \ram to \apm involves flattening the \ram program (represented as a DAG) into the \apm program (represented as a sequential list of instructions).
This flattening is implemented via a recursive \ram-to-\apm function $\texttt{compile} :: \text{\ram} \to [\text{instr}] \times [\text{reg}]$, a function that compiles a RAM expression into a sequence of instructions and returns the registers the result table is stored in.
For the complete definition of \texttt{compile}, see the Appendix.
Translation proceeds in the presence of a translation context $F_T$, also known as the EDB, which contains schema necessary for applying the translations and the provenance for using the proper tag operations.
We now examine two of these translation rules in detail to give examples of why the translation to \apm is challenging but makes the resulting programs amenable to GPU execution.

\paragraph{Project}
Projection is an example of the simplest parallelism \system exploits: row-level parallelism.
A projection expression $\pi_\alpha(\epsilon)$ consists of a projection expression $\alpha$ evaluated on an input table.
Critically, the expression is evaluated identically and without coordination across each fact of the input relation: this is a perfect fit for the SIMD paradigm employed by GPUs.
Propagating semiring tags is also straightforward: the provenance of each fact in the output is tied to exactly one fact in the input, so tags can be copied through to the result without modification.
Finally, performing allocation of the input facts is also straightforward, as the size of the output relation is identical to that of the input relation.
A concrete usage of lowering \texttt{project} to \apm can be seen in Figure~\ref{fig:compilation-example}, which features the compilation of a simple permutation projection.

\paragraph{Join}
Potentially the most important relational operator, join ($\bowtie_n$) forms the computational core of most \system programs, so it is important to find an efficient implementation.
Unfortunately, it is also more challenging than \texttt{project} for two reasons: (1) whereas each input fact in project produces exactly one output fact, with join each input fact can compute zero or more output facts and (2) rather than evaluating an expression against each row, join requires a membership test against the table being joined against.
To overcome these obstacles, \system takes inspiration from GPU hash-join algorithms \cite{shovon2023gpujoin}. 
\system has an additional requirement, however, to track provenance correctly in joins, where the provenance of each output fact is the product of the provenance of the input facts.
A concrete usage of lowering \texttt{join} to \apm can be seen in \autoref{fig:compilation-example}, which features the compilation of a join over two binary relations.

\subsection{Evaluating \apm}
\label{sec:apm-evaluation}

Once an \apm program is compiled, it is executed in a least fix-point iteration manner.
Note that the program executes continuously, updating the database each time, until no new facts are discovered.
To make this process efficient, \system employs a semi-naive evaluation strategy, which is a variant of the traditional naive evaluation strategy that avoids redundant computation.

Succinctly, semi-naive evaluation involves tracking a frontier of recently discovered facts, and only applying rules to frontier facts. 
This avoids the redundant computation of applying rules to stale facts that are known a priori to not produce new facts. 
Concretely, the database is partitioned into three sets of facts: delta facts (those that are computed during the current iteration), recent facts (those that were computed in the previous iteration), and stable facts (all other facts). 
After each iteration, the recent facts are merged with the stable facts and the delta facts become the recent facts.
Importantly, these semantics are codified in translation rules (see Appendix), which express semi-naive evaluation in terms of \apm instructions.
This means the deduplication of facts and the tracking of recent and stable facts is parallelized on the GPU just like the rest of the computation.

\subsection{Provenance Semiring Framework}
\label{sec:provenance-semiring}

As discussed prior, \system employs a GPU-accelerated provenance semiring framework with 7 implemented semirings covering discrete, probabilistic, and differentiable reasoning.
Specifically, \system supports unit, max-min-prob, add-mult-prob, top-$1$-proof, and the differentiable versions of the probabilistic semirings.
Tags in \apm are stored as an additional register alongside the column registers.
Since the tags may store boolean, floating point, and even complex data structures like dual-numbers and boolean formulae, we need each provenance to specify a fixed size for each tag.

\paragraph{Limitations}
Notably, Lobster departs from prior work by not supporting the fully general top-$k$-proof provenance~\cite{huang2021scallop}, but just the special case of top-$1$-proof.
This special case tracks just one conjunction of boolean variables for each fact, encoding that fact's proof set.
During disjunction, the provenance picks the more likely of the two proofs by computing the probabilities for each.
For conjunction, the provenance merges the two proofs and ensures that no conflict is present.
We find that this special case is sufficient for most practical applications and is much more efficient to compute than the general case.
Note that in this formulation, the size limit for a proof needs to be specified ahead of time.
We set the limit to 300 which is sufficient for all evaluated benchmarks.

\section{Optimizations}
\label{sec:optimizations}

Section~\ref{sec:language-compiler} describes the \apm language and its value as a principled way to handle Datalog with provenance on GPUs;
however, \apm shows additional utility as a platform for optimizations.
Here, we discuss optimizations from prior works that are easily implementable as \apm transformations, as well as novel optimizations enabled by the \apm runtime.

\subsection{Buffer Reuse and Management}
\label{subsec:buffer-reuse}
Regardless of evaluation strategy, query evaluation produces lots of temporary data, making allocation performance important.
Accordingly, every allocation in an \apm program is identified by an \texttt{alloc} instruction.
Due to the lack of branching and looping constructs, each iteration through an \apm program produces a fixed number of allocations. This enables two optimizations: arena allocation and buffer reuse.

\paragraph{Arena Allocation}
All data in registers is discarded after each iteration of an \apm program.
This fixed lifetime guarantee makes arena allocation~\cite{hanson1990fast} a good fit for \apm.
This observation makes allocations and deallocations in \apm essentially zero-cost, as allocation is reduced to bumping a pointer and deallocation is a no-op.

\paragraph{Buffer Reuse}
When arena allocation is impossible due to memory limits, allocation cost can still be amortized via buffer reuse.
Prior work on GPU Datalog execution made a similar discovery~\cite{sun2024moderndataloggpu}, noting that it was expensive to allocate a specific buffer required for merging relations after each iteration.
To address this, the authors over-allocate that buffer initially and reuse it across iterations.
Since each \apm buffer is identified by an \texttt{alloc} instruction, it is trivial to implement this optimization in \apm not just for a specific buffer, but for each buffer in the program.
This optimization is surprisingly effective, since the size of each register is strongly correlated with its size on the previous iteration.

\subsection{Hash Index Reuse via Static Registers}

Using a hash-based join is a boon for GPU acceleration, but building a hash index during each iteration of the fix-point loop negatively impacts performance.
To resolve this, we observe that frequently at least one input to a join is an EDB relation, which is constant across iterations.
Formally, Datalog programs are said to be ``linear recursive'' if each join has at most one IDB input, and we find that nearly all programs we consider are linear recursive.
In these cases, we can build a hash index during the first iteration of the fix-point loop and reuse it during successive iterations.

To realize this, we introduce the concept of static registers, modeled after the \texttt{static} keyword in C.
Static registers are initialized once and retain their values across iterations.
Marking the result of $\texttt{build}$ as $\texttt{static}$ when compiling \texttt{join} ensures that the hash index is reused across iterations.

\subsection{Batched Evaluation}
\label{subsec:batched}

\begin{figure}
    \centering
    \includegraphics[width=0.43\textwidth]{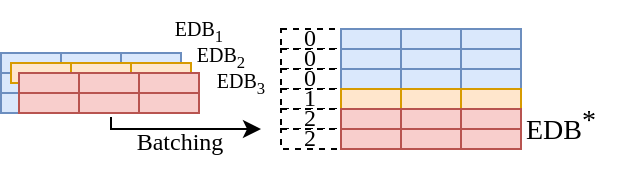}
    \caption{How \system incorporates batched input data via sample tagging. Distinct colors represent distinct samples within a batch.}
    \label{fig:batching}
\end{figure}

A key component of deep learning is grouping samples into \textit{batches} of samples that can be processed by the model in a single pass.
To truly integrate \system with deep learning as an end-to-end neurosymbolic tool, it should be aware of batching and able to process batches effectively.
Surprisingly, batching is a straightforward extension of the existing semantics.
Given a program $\overline{\phi}$ evaluated against a batch of three databases $F_1$, $F_2$, and $F_3$, we seek a database $F^*$ such that evaluating $\overline{\phi}$ against $F^*$ provides equivalent information to evaluating $\overline{\phi}$ against $F_1$, $F_2$, and $F_3$ separately.

We can construct $F^*$ without adding new constructs to \apm or \ram by taking our \apm columnar representation of a table and adding a new register $r_s$ to the front of the table that records the sample id.
Tables are now represented as a register pack $[r_s, r_1, \ldots, r_n, r_t]$, including the sample id, the columns, and the provenance semiring tags.
After execution, each IDB fact will have a sample tag which can be used to disambiguate results into per-sample databases that are returned to the user.
This is illustrated in Figure~\ref{fig:batching}.

Some desirable ramifications that naturally arise from this framing of batching are
(1) facts from separate batches cannot be joined together, so long as the width of each join operator is extended by one to include the batch tag;
(2) parallelizing over each element of the batch is implicit, as the runtime already parallelizes over the rows of a relation; and
(3) the additional memory footprint is minimal, since batches are seldom larger than 256 samples, meaning sample tags only take one byte of memory per row.

\section{Implementation}
\label{sec:implementation}

We build \system with a mixture of Rust, C++, and CUDA, reusing the front-end and query planner of Scallop to limit the scope of implementation. \system comprises approximately 2,000 new lines of Rust code and 9,000 new lines of CUDA/C++. We now discuss implementation details that fall outside the scope of the theory of \system's core compiler and runtime, yet are of practical interest and importance for implementing \system.

\subsection{Hash Table Design}
\label{sec:hash-table}
Crucial to \system's GPU-accelerated join algorithm is the existence of a lock-free, GPU hash table supporting parallel insertions and lookups.
While many implementations are possible, ours is inspired by previous work in \cite{shovon2023gpujoin}.
Namely, we adopt open-addressing with linear probing for collision resolution, enabling a contiguous memory representation with no indirection.
Unlike prior work, \system supports reasoning over relations with arbitrary width, so rather than  storing fact data directly in the hash table, we build hash \textit{indices} that map back to a row of the source table.
While this necessitates an additional random memory access to resolve collisions, it decouples the time and space complexity of the join from the width of the input relations.

\subsection{Bytecode Interpreter for Expression Evaluation}
\label{sec:bytecode}
Projection operations are pervasive in Datalog programs, yet generally account for a small portion of the runtime compared to operators like \texttt{join} due to projection's algorithmic simplicity.
Nonetheless, \system's optimizes the handling of \texttt{project} operations for GPU execution.
Specifically, there are two code paths for the implementation of \texttt{eval} in \apm.
Project expressions that permute or subset the columns of the relation can be evaluated as a series of columnar memory copies. Project expressions that contain arithmetic or comparison of tuple elements are compiled to bytecode for a simple stack machine, and each GPU thread executes this bytecode program against one fact with a small fixed-size stack residing in thread-local memory.

\subsection{Scheduling Stratum Offloading}
\label{subsec:scheduling-stratum}
\system relations start their life in CPU memory and once in GPU memory it is advantageous to continue operating on them with the GPU. To avoid the impact of high-latency CPU-GPU memory transfers, \system adopts an intelligent strategy for scheduling data transfers.

\system's begins by transferring facts to the GPU before the longest-running stratum (identified via a heuristic based on counting recursive joins) and back to the CPU after that stratum. From that longest-running stratum, we expand forwards and backwards in the static data-dependency graph to encompass adjacent strata as well, until the size of the stratum's inputs and outputs is small. Adopting a min-cut-like approach to GPU scheduling avoids spending excessive time in CPU-GPU transfers.

\subsection{Other Forms of Parallelism}
The primary parallelism exploited by \system is within each relational operator, and there is no parallelism in the execution of separate relational operators (they are executed sequentially). While the idea of parallelizing or pipelining operators is appealing, since \system's design requires very little CPU-GPU data movement, there is no opportunity to overlap data transfer with computation, so no performance improvement is achieved.

\section{Evaluation}
\label{sec:evaluation}

We empirically evaluate \system with the goal of demonstrating how well it performs in both training and inference tasks, and for the latter we explore a range of benchmarks that require differentiable, probabilistic, or discrete reasoning.

In the following sections, we introduce the benchmark tasks (\autoref{sec:benchmarks}) and the chosen baselines (\autoref{sec:baselines}) and present results in Sections \ref{sec:eval-training} and \ref{sec:eval-inference}.
All benchmarks are run on a machine with two 20-core Intel Xeon CPUs, a GeForce RTX 2080 Ti GPU, and 768 GB RAM,  with the exception of the discrete benchmarks, which have higher VRAM requirements and were run on a machine with two 24-core Intel Xeon CPUs, 1.5 TB RAM, and an NVIDIA A100 GPU with 80 GB VRAM.

\subsection{Benchmarks}
\label{sec:benchmarks}

We evaluate \system on a suite of ten benchmark tasks summarized in Table \ref{tab:tasks}.
Since \system is built on a flexible framework of provenance semirings, it supports differentiable, probabilistic, and discrete reasoning.
Correspondingly, we pick tasks across each of these reasoning modes to better illustrate the tradeoffs inherent in providing this flexibility.
The tasks span diverse application domains: natural language processing (CLUTRR), image processing (Pathfinder and HWF), program reasoning (Probabilistic Static Analysis), bioinformatics (RNA SSP), planning (PacMan-Maze), and graph databases (Transitive Closure, Same Generation, and CSPA).

\begin{table*}[h]
    \footnotesize
    \caption{Characteristics of benchmark tasks.}
    \vspace{-5px}
    \begin{tabular}{l|l|l|l|l|l}
    \toprule
    \multicolumn{1}{c|}{Task}     & Input  & \multicolumn{1}{c|}{Logic Program}                              & Kind   &\#Rules & Provenance        \\ \hline
    Pathfinder                    & Image  & Check if two dots are connected by a sequence of dashes.        & Diff.  & 2      & diff-top-1-proofs \\
    PacMan-Maze                   & Image  & Plan optimal next step by finding safe path from actor to goal. & Diff.  &  14    & diff-top-1-proofs \\
    HWF                           & Images & Parse and evaluate formula over recognized symbols.             & Diff.  &  13    & diff-top-1-proofs \\
    CLUTTR                        & Text   & Deduce kinship by recursively applying composition rules.       & Diff.  &   3    & diff-top-1-proofs \\
    Prob.~Static Analysis         & Code   &  Compute alarms with severity via probabilistic static analysis.& Prob.  &    28  & minmaxprob        \\
    RNA SSP                       & RNA    &  Parse an RNA sequence according to a context-free grammar.     & Prob.  &   28   & prob-top-1-proofs \\
    Transitive Closure            & Graph  &  Compute transitive closure of a directed graph.                & Disc.  &    2   & unit              \\
    Same Generation               & Graph  &  Compute graph vertices that are in the ``same generation''.    & Disc.  &    2   & unit              \\
    CSPA                          & Graph  &  A context sensitive pointer analysis.                          & Disc.  &    10  & unit              \\
    \bottomrule
    \end{tabular}
    \vspace{10px}
    \label{tab:tasks}
\end{table*}

The table describes each task's input, the functionality of the logic program, the kind of reasoning involved, and the number of rules.
The tasks requiring differentiable reasoning are taken from Scallop's evaluation (although we omit some tasks that do not have an obvious notion of scalability), the probabilistic reasoning tasks are crafted by us inspired by problems from the literature~\cite{st2024toward,sloma2016archiveii}, and the discrete reasoning tasks mirror the evaluation of the latest work in this space, FVLog \cite{sun2025columnorienteddataloggpu}.

Notably, our results focus exclusively on performance, with no mention of correctness. This is because in all cases each system under test produces identical results: for differentiable tasks, the model reaches identical accuracy whether Scallop or \system is used, and for probabilistic and discrete tasks, the logic programs are identical and therefore produce identical results. We refer curious readers who desire a more thorough discussion of comparing accuracy between pure-neural and neurosymbolic models to the Scallop paper~\cite{scallop}, the results of which we partially replicate here with \system.

We next briefly describe each of the tasks.

\textbf{Pathfinder} This task is discussed in-depth in Section~\ref{sec:motivating} and requires reasoning over an image to determine if two dots are connected by a sequence of dashes.

\textbf{PacMan-Maze}
In this task, a neurosymbolic reinforcement learning agent aims to solve a 2D maze given only an image of the maze.
The neural portion executes a CNN to predict enemy locations in the maze, and the symbolic portion plans a safe path to the goal.
Our formulation leverages curriculum learning: the agent first learns in a 5-by-5 maze and then moves to a 20-by-20 grid.

\textbf{HWF}
The Handwritten Formula (HWF) \cite{li2020closed} task requires parsing and evaluating a formula of handwritten digits and operators, given supervision only on the final value.
The dataset consists of formulas of varying length, meaning naive parallelism strategies like processing each formula in a batch separately will fall short due to work imbalances. Further, the symbolic program requires support for floating-point data and floating-point arithmetic operations.

\textbf{CLUTRR}
CLUTRR is a natural language reasoning task about family kinship relations \cite{sinha2019clutrr}.
The input contains a natural language passage about a family with each sentence in the passage hinting at kinship relations.
The goal is to infer the relationship between a given pair of characters; however, the target relation is not stated explicitly in the passage and it must be deduced through a reasoning chain.
The most difficult problem in the evaluation dataset requires reasoning through a chain of length 10.

\textbf{Probabilistic Static Analysis}
This benchmark extends static program analysis with probabilistic inputs.
Specifically, analysis inputs are annotated with probabilities to reflect the system's confidence.
These probabilities are propagated to the output and used to rank results in order to decrease the visibility of false positives \cite{st2024toward}.

\textbf{RNA SSP}
This task performs RNA Secondary Structure Prediction (SSP) using the ArchiveII~\cite{sloma2016archiveii} dataset.
RNA SSP discovery is of widespread interest in the medical community, as the secondary structure of RNA molecules is crucial for understanding their function.
Our neurosymbolic solution uses a Datalog program to parse an RNA sequence according to a context-free grammar, given probabilistic input from a transformer model. The dataset consists of a set of 475 RNA sequences of length 28 to 175.

\textbf{Transitive Closure}
This benchmark computes the reachability of nodes in a graph using discrete reasoning. We use graphs from the SNAP graph repository~\cite{snapnets} that take at least 1 second to process.

\textbf{Same Generation}
This benchmark computes which nodes in a directed graph are the same distance from a common ancestor, i.e. which nodes are in the ``same generation''. We use the same graphs as in the Transitive Closure benchmark.

\textbf{Context Sensitive Pointer Analysis (CSPA)}
This benchmark computes a context sensitive pointer analysis for a program. We mirror the evaluation of GDLog~\cite{sun2024moderndataloggpu} both in the Datalog program that we use and the input graphs.

\subsection{Baselines}
\label{sec:baselines}

We compare \system to several other systems, as shown previously in \autoref{fig:overview}. While no prior system matches \system's feature set, comparisons to more limited systems help us gauge \system's performance across a range of use-cases.

\textbf{Scallop} supports differentiable reasoning with provenance semirings like \system, but supports only batch-level CPU multicore parallelism and therefore struggles to scale with problem and data complexity. We do not evaluate on DeepProbLog~\cite{manhaeve2021deepproblog}, a similar CPU-only system supporting differentiable and probabilistic reasoning, as previous work has shown that Scallop's performance is uniformly superior~\cite{scallop}.

\textbf{ProbLog} \cite{anton2015problog2} provides discrete and probabilistic reasoning, but does not support GPU acceleration or CPU multi-threading.
Notably, ProbLog executes logic programs via stable model semantics, indicating it may find certain programs easier or harder than other baselines, which are all based on bottom-up search.
Additionally, while Problog has been used to study approximate probabilistic reasoning~\cite{renkens2012approximateproblog}, approximate inference is not implemented in the publicly released Problog. Therefore, in our experiments Problog performs exact inference, as opposed to the more scalable approximate inference that \system and Scallop perform.

\textbf{FVLog} \cite{sun2025columnorienteddataloggpu} supports only discrete reasoning, but does leverage GPU acceleration.
FVLog is specialized for a different sort of workload than \system: it targets large batch analysis jobs that may span minutes, whereas \system emphasizes running the same program multiple times as a component of a neurosymbolic model.
Notably, FVLog does not offer a Datalog front-end and query planner, meaning that all FVLog programs are human-written, low-level, relational algebra programs.
As FVLog is faster than prior GPU discrete Datalog systems (like GDLog \cite{sun2024moderndataloggpu}), we compare only to FVLog for brevity.

\textbf{\souffle} \cite{bernhard2016souffle} is a state-of-the-art, multicore, CPU Datalog engine.
While it does not support differentiable or probabilistic reasoning, comparing with \souffle helps reveal the benefits of GPU acceleration versus CPU optimizations.

\subsection{\system for Training}
\label{sec:eval-training}

\begin{figure}[t]
    \centering
    \scriptsize
    \begin{minipage}{0.47\textwidth}
        \pgfplotstableread[col sep=comma]{training.txt}{\psaData}
        \begin{tikzpicture}
            \begin{axis}[
                width =\textwidth,
                height= 3.6cm,
                ybar=2*\pgflinewidth,
                nodes near coords,
                every node near coord/.append style={font=\small},
                bar width=5mm,
                ymajorgrids = true,
                ylabel = {Speedup over Scallop ($\times$)},
                symbolic x coords={CLUTTR,HWF,Pathfinder,Pacman},
                xtick = data,
                x tick label style={},
                scaled y ticks = false,
                ymin=0,
                ymax=20,
                legend cell align=left,
                legend columns=3,
                legend image code/.code={
                    \draw[#1] (0cm,-0.1cm) rectangle (0.3cm,0.1cm);
                },
                legend style={
                        at={(0.5,0.90)},
                        anchor=north,
                        column sep=1ex
                }
            ]
                \addplot[style={mydeepblue,fill=myblue,mark=none}]
                    table [x={Name}, y={Ours}] 
                    {\psaData};
            \end{axis}
        \end{tikzpicture}
        \vspace{-7px}
        \captionof{figure}{\system's speedup over Scallop on training tasks.}
        \label{fig:training-results}
    \end{minipage}
\end{figure}

To evaluate the extent to which \system improves performance in the training pipeline, we compare the total training time of \system against Scallop, the only other system that supports neurosymbolic training.
For each task, training is run until convergence rather than a pre-determined number of epochs, and therefore takes a task-specific number of epochs.
However, for a given task, both \system and Scallop take the same number of epochs, as they are executing the same Datalog program and produce identical results.

The results in \autoref{fig:training-results} reveal that \system can achieve significant speedups in end-to-end training time compared to Scallop, ranging from 1.2x to 16x.
\autoref{fig:training-results} includes the cost of neural computations, which are already heavily optimized on GPU hardware via Pytorch \cite{paszke2019pytorch} and unaffected by \system, so Amdahl's Law limits the potential end-to-end speedup. Pacman is an exception as it performs extensive symbolic computation which \system can greatly accelerate.

\subsection{\system for Inference}
\label{sec:eval-inference}

Beyond training of neurosymbolic models, we also evaluate \system's performance on a range of neurosymbolic inference tasks.
In neurosymbolic inference, the neural component is pre-trained.
For example, with Pacman a pre-trained neural classifier identifies in-game objects, but the symbolic program still needs to be run to determine the path to the goal for each game board.
We report average times across a set of samples: CLUTRR processes relationship graphs from 13 text passages, HWF evaluates 160 formulas of length 13, Pathfinder is evaluated on a set of 1216 images, and PacMan-Maze solves 50 mazes on a 15x15 grid.

\autoref{fig:nesy-inference-results} shows that \system can obtain significant speedups over Scallop. Compared to results from training (\autoref{fig:training-results}), during inference benchmarks like CLUTTR and Pathfinder spend significantly less time in neural computation which leads to larger speedups with \system. Notably, the speedups for Pacman is less for inference than for training. We believe this is because the trained model solves each maze in fewer steps, making the symbolic computation a smaller fraction of the total runtime.

Next we explore how well \system scales to larger problem sizes. We choose the benchmarks that can be scaled most naturally: for Pathfinder we increase the resolution of the analysis and for Pacman the maze size. We choose Scallop as the baseline as it is the only system that supports these neurosymbolic workloads. Further, we only consider the symbolic computation time to more precisely measure \system's improvement. In \autoref{fig:nesy-inference-scaling} we see that \system handles large problem sizes better than Scallop, although the speedup plateaus as the problem size becomes large enough to saturate GPU memory bandwidth. \autoref{fig:nesy-inference-scaling} also shows an ablation study that justifies some of the optimizations in \system---without allocation optimization and stratum scheduling, performance rapidly degrades to near equal with Scallop on problem sizes greater than 20.

\begin{figure}[t]
    \centering
    \scriptsize
    \begin{minipage}{0.47\textwidth}
        \pgfplotstableread[col sep=comma]{nesy-inference.txt}{\psaData}
        \begin{tikzpicture}
            \begin{axis}[
                width =\textwidth,
                height= 3.6cm,
                ybar=2*\pgflinewidth,
                nodes near coords,
                every node near coord/.append style={font=\small},
                bar width=5mm,
                ymajorgrids = true,
                ylabel = {Speedup over Scallop ($\times$)},
                symbolic x coords={CLUTTR,HWF,Pathfinder,Pacman},
                xtick = data,
                x tick label style={yshift=1.5mm, rotate=-17},
                scaled y ticks = false,
                ymin=0,
                ymax=5,
                ytick distance = 1,
                legend cell align=left,
                legend columns=3,
                legend image code/.code={
                    \draw[#1] (0cm,-0.1cm) rectangle (0.3cm,0.1cm);
                },
                legend style={
                        at={(0.5,0.90)},
                        anchor=north,
                        column sep=1ex
                }
            ]
                \addplot[style={mydeepblue,fill=myblue,mark=none}]
                    table [x={Name}, y={Ours}] 
                    {\psaData};
            \end{axis}
        \end{tikzpicture}
        \vspace{-5px}
        \captionof{figure}{\system's speedup over Scallop on neurosymbolic inference tasks.}
        \label{fig:nesy-inference-results}
    \end{minipage}
\end{figure}

\begin{figure}

\definecolor{red1}{RGB}{200,55,40}
\definecolor{red2}{RGB}{255,140,0}
\definecolor{orange1}{RGB}{255,140,0}
\definecolor{pink1}{RGB}{255,105,180}
\definecolor{blue1}{RGB}{107,174,214}
\definecolor{blue2}{RGB}{49,130,189}

\begin{minipage}[b]{0.48\textwidth}
    \begin{subfigure}{\textwidth}
        \centering
        \scriptsize
        \begin{tikzpicture}
            \pgfplotstableread[col sep=comma]{pacman-scaling.dat}{\pacmanData}
            \begin{axis}[
                height=4.0cm,
                width=\textwidth,
                enlargelimits=false,
                xlabel=Grid Size,
                ylabel={Speedup over Scallop ($\times$)},
                enlarge x limits=0.05,
                enlarge y limits=0.1,
                ytick distance=1,
                x label style={at={(axis description cs:0.5,-0.15)},anchor=north},
                y label style={at={(axis description cs:-0.05,.5)},anchor=south},
                legend columns=2,
                legend style={font=\tiny,at={(0.15,0.95)},anchor=north}
            ]
            
            \addplot [only marks,mark size=1pt,red1]  table [x={size}, y expr=\thisrow{scallop}/\thisrow{nono}] {\pacmanData};
            \addplot [only marks,mark size=1pt,orange1]  table [x={size}, y expr=\thisrow{scallop}/\thisrow{noyes}] {\pacmanData};
            \addplot [only marks,mark size=1pt,pink1]  table [x={size}, y expr=\thisrow{scallop}/\thisrow{yesno}] {\pacmanData};
            \addplot [only marks,mark size=1pt,blue1]  table [x={size}, y expr=\thisrow{scallop}/\thisrow{yesyes}] {\pacmanData};

            \addplot [red1] table [x={size}, y expr=\thisrow{scallop}/\thisrow{nono}] {\pacmanData};
            \addplot [orange1] table [x={size}, y expr=\thisrow{scallop}/\thisrow{noyes}] {\pacmanData};
            \addplot [pink1] table [x={size}, y expr=\thisrow{scallop}/\thisrow{yesno}] {\pacmanData};
            \addplot [blue1] table [x={size}, y expr=\thisrow{scallop}/\thisrow{yesyes}] {\pacmanData};
            \legend{None,Stratum,Alloc,Both}
            \end{axis}
        \end{tikzpicture}
        \vspace{-5px}
        \caption{\system's scalability on Pacman.}
    \end{subfigure}
    \begin{subfigure}{\textwidth}
        \centering
        \scriptsize
        \begin{tikzpicture}
            \pgfplotstableread[col sep=comma]{pathfinder-scaling.dat}{\pathfinderData}
            \begin{axis}[
                height=4.2cm,
                width=\textwidth,
                enlargelimits=false,
                xlabel=Grid Size,
                ylabel={Speedup over Scallop ($\times$)},
                enlarge x limits=0.05,
                enlarge y limits=0.1,
                ytick distance=1,
                x label style={at={(axis description cs:0.5,-0.15)},anchor=north},
                y label style={at={(axis description cs:-0.05,.5)},anchor=south},
                legend columns=2,
                legend style={font=\tiny,at={(0.15,0.95)},anchor=north}
            ]
            
            \addplot [only marks,mark size=1pt,red1]  table [x={size}, y expr=\thisrow{scallop}/\thisrow{nono}] {\pathfinderData};
            \addplot [only marks,mark size=1pt,orange1]  table [x={size}, y expr=\thisrow{scallop}/\thisrow{noyes}] {\pathfinderData};
            \addplot [only marks,mark size=1pt,pink1]  table [x={size}, y expr=\thisrow{scallop}/\thisrow{yesno}] {\pathfinderData};
            \addplot [only marks,mark size=1pt,blue1]  table [x={size}, y expr=\thisrow{scallop}/\thisrow{yesyes}] {\pathfinderData};

            \addplot [red1] table [x={size}, y expr=\thisrow{scallop}/\thisrow{nono}] {\pathfinderData};
            \addplot [orange1] table [x={size}, y expr=\thisrow{scallop}/\thisrow{noyes}] {\pathfinderData};
            \addplot [pink1] table [x={size}, y expr=\thisrow{scallop}/\thisrow{yesno}] {\pathfinderData};
            \addplot [blue1] table [x={size}, y expr=\thisrow{scallop}/\thisrow{yesyes}] {\pathfinderData};
            \legend{None,Stratum,Alloc,Both}
            \end{axis}
        \end{tikzpicture}
        \vspace{-5px}
        \caption{\system's scalability on Pathfinder.}
    \end{subfigure}
    \caption{\system's scalability on Pathfinder and Pacman in the presence of various optimizations. ``None'' indicates no optimizations, ``Stratum'' includes the stratum scheduling heuristic of \autoref{subsec:scheduling-stratum}, ``Alloc'' includes the allocation optimizations of \autoref{subsec:buffer-reuse}, and ``Both'' includes  both optimizations.}
    \label{fig:nesy-inference-scaling}
\end{minipage}
\end{figure}

Beyond neurosymbolic workloads, we also run two \textbf{probabilistic} inference workloads: Probabilistic Static Analysis (PSA) and RNA Secondary Structure Prediction (RNA SSP). As these two workloads require only probabilistic (not differentiable) reasoning, we attempted to run them with ProbLog \cite{anton2015problog2}. However, all of the PSA and RNA SSP runs hit our 2-hour timeout, except for PSA on sunflow-core which was 60\% slower than Scallop and 30x slower than \system. As stated in \autoref{sec:baselines}, we believe this is explained by ProbLog's exact probabilistic inference.

\autoref{fig:psa-results} shows that, on the PSA benchmark, \system again offers significant speedups over Scallop when performing static analysis across a range of source programs. With RNA SSP (\autoref{fig:rna-results}), on the very shortest sequence (28 base pairs) \system is 40\% slower than Scallop. However, on all of the other sequences \system achieves a speedup, frequently by two orders of magnitude. The speedup correlates with sequence length, making \system even more valuable on longer sequences which are, generally, of greater biological interest.

\begin{figure}[t]
    \centering
    \scriptsize
    \begin{minipage}{0.47\textwidth}
        \pgfplotstableread[col sep=comma]{psa-speedup.txt}{\psaData}
        \begin{tikzpicture}
            \begin{axis}[
                width =\textwidth,
                height= 3.6cm,
                ybar=2*\pgflinewidth,
                nodes near coords,
                every node near coord/.append style={font=\small},
                bar width=5mm,
                ymajorgrids = true,
                ylabel = {Speedup over Scallop ($\times$)},
                symbolic x coords={avrora,biojava,graphchi,jme3,pmd,sunflow-core,sunflow},
                xtick = data,
                x tick label style={yshift=1.5mm, rotate=-17},
                scaled y ticks = false,
                ymin=0,
                ymax=22,
                legend cell align=left,
                legend columns=3,
                legend image code/.code={
                    \draw[#1] (0cm,-0.1cm) rectangle (0.3cm,0.1cm);
                },
                legend style={
                        at={(0.55,0.90)},
                        anchor=north,
                        column sep=1ex
                }
            ]
                \addplot[style={mydeepblue,fill=myblue,mark=none}]
                    table [x={Name}, y={Ours}] 
                    {\psaData};
            \end{axis}
        \end{tikzpicture}
        \vspace{-5px}
        \captionof{figure}{\system's speedup over Scallop on Probabilistic Static Analysis.}
        \label{fig:psa-results}
    \end{minipage}
\end{figure}

\begin{figure}[t]
    \centering
    \footnotesize
    \begin{subfigure}{0.48\textwidth}
            \centering
            \scriptsize
            \begin{tikzpicture}
                \definecolor{myblue}{HTML}{4a86e8}
                \pgfplotstableread[col sep=tab]{rna-times.txt}{\rnaTimesData}
                \begin{axis}[
                    height=3.8cm,
                    width=\textwidth,
                    enlargelimits=false,
                    xlabel=RNA Sequence Length,
                    ylabel=Speedup over Scallop ($\times$),
                    ytick distance=100,
                    ymin=1,
                    ymax=600,
                    x label style={at={(axis description cs:0.5,-0.15)},anchor=north},
                    y label style={at={(axis description cs:-0.08,.5)},anchor=south},
                ]
                \addplot [only marks,mark size=0.7pt,myblue] 
                    table [x={Length}, y expr=\thisrow{CPU}/\thisrow{GPU}] 
                    {\rnaTimesData};
                
                \end{axis}
            \end{tikzpicture}
    \end{subfigure}
    \vspace{-5px}
    \caption{
        \system speedup over Scallop on RNA SSP.
    }
    \label{fig:rna-results}
\end{figure}

\begin{figure}
    \begin{minipage}{0.48\textwidth}
        \scriptsize
        \pgfplotstableread[col sep=comma]{transitive-closure-speedup.txt}{\tcData}
        \begin{tikzpicture}
            \begin{axis}[
                width =1.06\textwidth,
                height= 3.6cm,
                ybar=2*\pgflinewidth,
                bar width=1.2mm,
                ymajorgrids = true,
                ylabel = {Speedup over \souffle ($\times$)},
                symbolic x coords={com-dblp,Gnu31,p2p-Gnu24,cit-HepPh,cit-HepTh,fe-body,fe-sphere,loc-Brightkite,p2p-Gnu25,p2p-Gnu30,SF.cedge,usroad,vsp-finan},
                xtick = data,
                x tick label style={yshift=1.5mm, rotate=-25},
                scaled y ticks = false,
                ymax=80,
                legend columns=1,
                legend cell align=left,
                legend image code/.code={
                    \draw[#1] (0cm,-0.1cm) rectangle (0.3cm,0.1cm);
                },
                legend style={
                        anchor=north east,
                        column sep=1ex
                }
            ]
                \addplot[style={mydeepblue,fill=myblue,mark=none}]
                    table [x={Name}, y={Lobster}] 
                    {\tcData};
                \addplot[style={mydeeporange,fill=myorange,mark=none}]
                     table [x={Name}, y={FVLog}] 
                    {\tcData};
                \legend{\system,FVLog}
            \end{axis}
        \end{tikzpicture}
        \vspace{-10px}
        \captionof{figure}{Speedup over \souffle on Transitive Closure.}
        \label{fig:tc-results}
    \end{minipage}
\end{figure}

\begin{table}[h]
    \caption{The runtime of \system versus FVLog on the Same Generation task. ``OOM'' indicates that the system ran out of memory.}
    \vspace{-5px}
    \begin{tabular}{lcc}
    \toprule
    Dataset           & \system (s)  & FVLog (s) \\ \hline
    fe-sphere         &  \textbf{3.91 } & 12.99 \\
    CA-HepTH          &  \textbf{2.16 } & 6.40  \\
    ego-Facebook      &  \textbf{0.53 } & OOM   \\
    Gnu31             &   OOM           & OOM   \\
    fe\_body          &  \textbf{10.17} & 21.17 \\
    loc-Brightkite    &  \textbf{1.45 } & OOM   \\
    SF.cedge          &  \textbf{14.01} & 23.72 \\
    com-dblp          &   OOM           & OOM   \\
    usroad            &   OOM           & OOM   \\
    fc\_ocean         &  \textbf{2.17 } & 4.67  \\
    vsp\_finan        &  OOM & \textbf{90.10} \\
    \bottomrule
    \end{tabular}
    \vspace{10px}
    \label{tab:sg}
\end{table}

\begin{table}[h]
    \caption{The runtime of \system versus FVLog on the CSPA task.}
    \vspace{-5px}
    \begin{tabular}{lcc}
    \toprule
    Dataset        & \system (s)  & FVLog (s) \\ \hline
    httpd        &  3.61  & \textbf{2.57} \\
    linux        &  \textbf{1.81}  & 3.91 \\
    postgres     &  \textbf{3.32}  & 4.39 \\
    \bottomrule
    \end{tabular}
    \vspace{10px}
    \label{tab:cspa}
\end{table}

Finally, we examine \system's performance on \textbf{discrete reasoning} tasks that require neither differentiable nor probabilistic reasoning. We first run the Transitive Closure benchmark from FVLog \cite{sun2025columnorienteddataloggpu} on a range of input graphs \cite{snapnets}. For these inputs, ProbLog always hit our 2-hour timeout, and Scallop (when not timing out) has a 30-90x slowdown over \souffle, so we omit results for those systems. \autoref{fig:tc-results} shows results for \system and FVLog and demonstrates that, despite \system's generality, it offers competitive performance even against more specialized systems, consistently beating the CPU-only \souffle and often surpassing the GPU-accelerated FVLog. We attribute this to \system's adoption of APM. FVLog in particular lacks any IR and thus forgoes the opportunities afforded by IR-level optimizations (\autoref{sec:optimizations}).

Since FVLog is the sole competitive system, we only run the remaining two discrete tasks, Same Generation and CSPA, on \system and FVLog, displaying the results in \autoref{tab:sg} and \autoref{tab:cspa} respectively. We observe that for the Same Generation task, \system is at least twice as fast on each dataset and that there are multiple datasets that \system processes which FVLog runs out of memory on. The one exception is the vsp\_finan dataset, where \system runs out of memory while FVLog finishes in 90 seconds. We believe this is due to the fact that \system is more general and therefore requires more memory to store intermediate results. For the CSPA task, \system and FVLog are approximately matched, with \system exhibiting a geometric mean speed up of 1.27x over FVLog.

\section{Related Work}
\label{sec:related}

While there is a wealth of work on GPU-acceleration for SQL in both research and industry (e.g., \cite{pgstrom, heavydb}), we focus our related work discussion on systems for logic programming beyond SQL.
We relate \system to works along three directions: accelerated Datalog engines, probabilistic and differentiable programming, and neurosymbolic methods.

\textbf{High-Performance Datalog} 
A variety of Datalog-based systems have been built for program analysis \cite{bernhard2016souffle,souffle-progam-analysis,recstep} and even enterprise database applications \cite{molham2015logicblox}, though these systems run exclusively on the CPU. 
The FVLog system \cite{shovon2023gpujoin,sun2024moderndataloggpu} provides a Datalog engine implemented for GPUs, but it lacks support for the probabilistic and differentiable reasoning needed for deep learning integration. 
Moreover, FVLog focuses on the domain of large analytics queries, which emphasizes simpler queries executed against large databases, which is not a focus for \system.
Notably, FVLog does not come with a query planner and user-facing front-end, meaning that users need to directly interact with low-level relational algebra operations supported by the system.

\textbf{Probabilistic and Differentiable Programming} 
\textit{Probabilistic programming} allows programmers to model distributions and perform probabilistic sampling and inference \cite{anton2015problog2,bingham2018pyro,ge2018t,meent2018intro-ppl}.
\textit{Differentiable programming} systems allow programmers to write code that is differentiable and therefore amenable to use during neural network training. Symbolic and automatic differentiation \cite{bayd15autodiff} are commonly used in popular ML frameworks such as PyTorch and others~\cite{paszke2019pytorch, abadi2015tensorflow, frostig2018jax}.

Probabilistic programs are not in general differentiable and thus cannot be run during training. The differentiable programming systems described above are designed for real-valued functions and are not compatible with logic programming.
\system, on the contrary, focuses on the differentiability of logic programs with probabilities.

\textbf{Neurosymbolic Methods} 
The emerging domain of neurosymbolic computation combines symbolic reasoning into existing data-driven learning systems. 
There have been a large number of successful neurosymbolic systems across a range of machine learning domains like computer vision and natural language processing \cite{yi2018nsvqa, mao2019nscl, li2020closed, wang2019satnet, xu2022dont, chen2020nerd, minervini2020ctp, xu18semanticloss, shah2020learning, chen2021webqa, cheng2022binding, zelikman2023parsel, cohen2017tensorlog, solko2024data, yang2023neurasp, chen2021neurallog, mao2019neuro, manhaeve2018deepproblog, manhaeve2021deepproblog}.
\system builds upon the Scallop neurosymbolic programming language \cite{huang2021scallop,li2023scallop}, as Scallop is general enough to implement other neurosymbolic systems \cite{mao2019nscl,chen2020nerd,xu18semanticloss,xu2022dont}. 
However, \system improves upon the CPU-only Scallop by using GPU acceleration to provide higher performance and the ability to scale to larger datasets, as we demonstrated in Section~\ref{sec:evaluation}.

\section{Conclusion}
\label{sec:conclusion}

We have described the design and implementation of the \system neurosymbolic engine. With existing engines, symbolic computation can quickly become the bottleneck when neural computations benefit from domain-specific hardware accelerators like GPUs. \system shows how Datalog programs can also take advantage of GPUs, providing large speedups and strong scalability over CPU-only engines like Scallop.

\begin{acks}
We would like to thank our shepherd, Professor Michael O'Boyle. This paper is based upon work supported by the National Science Foundation under Grant No. 2313010. Any opinions, findings, and conclusions or recommendations expressed in this material are those of the author and do not necessarily reflect the views of the National Science Foundation.
\end{acks}

\bibliographystyle{ACM-Reference-Format}
\balance
\bibliography{references}

% \newpage
\appendix
\section{\ram to \apm Translation}
We detail the \texttt{compile} function used to translate \ram to \apm in \autoref{fig:ram-apm-translation}.

\begin{figure*}
    \begin{subfigure}[b]{0.47\linewidth}
        \footnotesize
        \centering
        \bgroup
        \def\arraystretch{1.9}
        \begin{tabular}{c|l}
            \textbf{Compile Expression} & \textbf{Instructions} \\
            \hline
            \makecell{
                $\text{compile}(\pi_{\alpha_{n,m}}$$(\epsilon), D)$ 
                \textit{``Project''}
            } & \makecell[l]{
                $\textrm{\textbf{let} } ([s_1,\ldots,s_n,s_t], c) = \text{compile}(\epsilon,D) \textrm{ \textbf{in}} $\\
                ($c \cdot \{ $\\
                $[d_1,\ldots,d_m,d_t] \gets \instr{alloc}(\text{size}(s_1)) $\\
                $[d_1,\ldots,d_m] \gets \instr{eval}\langle\alpha_{n,m}\rangle ([s_1,\ldots,s_n]) $\\
                $d_t \gets \instr{copy}(s_t)$ \}, $[\overline{d_n}, d_t]$)
            } \\
            \hline
            \makecell{
                $\text{compile}(\rho(x_1,\ldots,$$x_n), D)$ 
                \textit{``Relation''}
            } & \makecell[l]{
                (\{\instr{alloc}([$s_1,\ldots,s_n,s_t]$, $\text{size}(D(\rho))$) \\
                $[\overline{s_n}, s_t]$ $\gets$ \instr{load}$\langle D(\rho) \rangle()$ \},
                $[\overline{s_n}, s_t]$)
            } \\
            \hline
            \makecell{
                $\text{compile}(p \gets \epsilon, D)$ 
                \textit{``Update''}
            } & \makecell[l]{
                $\textrm{\textbf{let} } ([s_1,\ldots,s_n,s_t], c) = \text{compile}(\epsilon) \textrm{ \textbf{in}} $\\
                $(c \cdot \{ $\\
                \instr{store}($p$, $[s_1,\ldots,s_n,s_t]$) \}, $\emptyset$)
            } \\
            \hline
            \makecell{
                $\text{compile}(p_1 \gets \epsilon_1,$ 
                $\ldots, p_n \gets \epsilon_n)$ \\
                \textit{``Stratum''}
            } & \makecell[l]{
                $\textrm{\textbf{let} } (\_, c_1) = \text{compile}(p_1 \gets \epsilon_1) \textrm{ \textbf{in}} $\\
                \ldots \\
                $\textrm{\textbf{let} } (\_, c_n) = \text{compile}(p_n \gets \epsilon_n) \textrm{ \textbf{in}} $\\
                ($c_1 \cdot \ldots \cdot c_n \cdot \{ $\\
                \instr{alloc}($\overline{s_n}, \text{size}(F_T^\text{stable})(\rho)$) \\
                $\overline{s_n} \gets \instr{load}\langle F_T^\text{stable}(\rho)\rangle()$ \\
                \instr{alloc}($\overline{r_n}, \text{size}(F_T^\text{recent})(\rho)$) \\
                $\overline{r_n} \gets \instr{load}\langle F_T^\text{recent}(\rho)\rangle()$ \\
                \instr{alloc}($\overline{d_n}, \text{size}(F_T^{\Delta}(\rho))$) \\
                $\overline{d_n} \gets \instr{load}\langle F_T^\Delta (\rho)\rangle()$ \\
                $\instr{alloc}(\overline{s_n^\text{new}}, \text{size}(\overline{s_n}+\overline{r_n}))$ \\
                $\overline{s_n^\text{new}} \gets \instr{merge}(\overline{s_n}, \overline{r_n})$ \\
                $\instr{alloc}([\overline{d_n^\text{sorted}}, \overline{d_n^\text{unique}}], \text{size}(\overline{d_n}))$ \\
                $\overline{d_n^\text{sorted}} \gets \instr{sort}(\overline{d_n})$ \\
                $\overline{d_n^\text{unique}} \gets \instr{unique}(\overline{d_n^\text{sorted}})$ \\
                $\instr{store}\langle F_T^\text{stable} (\rho)\rangle(\overline{s_n^\text{new}})$ \\
                $\instr{store}\langle F_T^\text{recent} (\rho)\rangle(\overline{d_n^\text{unique}})$ \\
                $\instr{store}\langle F_T^\Delta (\rho)\rangle(\emptyset)$
                \} \textbf{for} $\rho$ \textbf{in} unique($\rho_1,\ldots,\rho_n$) \\
                , $\emptyset$)
            } \\
            \hline
            \makecell{
                $\text{compile}(\epsilon_1 \bowtie_w \epsilon_2$$, D)$ \\
                \textit{``Join''}
            } &
            \makecell[l] {
                $\textrm{\textbf{let} }(\overline{w_n}, c_1) = \textrm{join\textsubscript{impl}}(\epsilon_1 \bowtie_w \epsilon_2, F_T^{\text{stable}}, F_T^{\text{recent}})$ in \\
                $\textrm{\textbf{let} }(\overline{x_n}, c_2) = \textrm{join\textsubscript{impl}}(\epsilon_1 \bowtie_w \epsilon_2, F_T^{\text{recent}}, F_T^{\text{stable}})$ in \\
                $\textrm{\textbf{let} }(\overline{y_n}, c_3) = \textrm{join\textsubscript{impl}}(\epsilon_1 \bowtie_w \epsilon_2, F_T^{\text{recent}}, F_T^{\text{recent}})$ in \\
                (\{
                \instr{alloc}($\overline{z_n}, \text{size}(w_1) + \text{size}(x_1) + \text{size}(y_1)$) \\
                $\overline{z_n} \gets \instr{append}(\overline{w_n}, \overline{x_n}, \overline{y_n})$ \}, $\overline{z_n}$)
            } \\
            \hline
            \makecell{
                $\text{join\textsubscript{impl}}(\epsilon_1 \bowtie_w$ $\epsilon_2, D_1, D_2)$ \\
            } &
            \makecell[l]{
                $\textrm{\textbf{let} } ([a_1,\ldots,a_n,a_t], c_1) = \text{compile}(\epsilon_1, D_1) \textrm{ \textbf{in}} $\\
                $\textrm{\textbf{let} } ([b_1,\ldots,b_m,b_t], c_2) = \text{compile}(\epsilon_2, D_2) \textrm{ \textbf{in}} $\\
                ($c_1 \cdot c_2 \cdot \{ $\\
                \instr{alloc}($h$, $\text{size}(a_1) * \mathcal{O}$) \\
                \instr{static} $h$ $\gets$ \instr{build}($[a_1,\ldots,a_w]$) \\
                \instr{alloc}([$c,o$], size($b_1$)) \\
                $c$ $\gets$ \instr{count}($[b_1,\ldots,b_w]$,h,$[a_1,\ldots,a_w]$) \\
                $o$ $\gets$ \instr{scan}($c$) \\
                \instr{alloc}([$i_l$, $i_r$, $d_1,\ldots,d_{n+m-w},d_t$],last(o)) \\
                $[i_l,i_r]$ $\gets$ \instr{join}$\langle w \rangle$($\overline{b_m}$, $\overline{a_n}$,$h$,$c$,$o$) \\
                $[d_1,\ldots,d_n]$ $\gets$ \instr{gather}($i_l$, $\overline{a_n}$) \\
                $[d_{n+1},\ldots,d_{n+m-w}]$ $\gets$ \instr{gather}($i_r$, $\overline{b_m}$) \\
                $d_t$ $\gets$ \instr{gather}$\langle \otimes \rangle$([$i_l$, $i_r$], $[a_t,b_t]$) \}, $[\overline{d_n}, d_t]$)
            }
            
        \end{tabular}
        \egroup
    \end{subfigure}
    \begin{subfigure}[b]{0.47\linewidth}
        \footnotesize
        \centering
        \bgroup
        \def\arraystretch{1.9}
        \begin{tabular}{c|l}
            
        \end{tabular}
        \egroup
    \end{subfigure}
    \caption{
    A subset of the function $\texttt{compile} :: \text{\ram} \to [\text{instr}] \times [\text{reg}]$ which translates a \ram program to \apm via a per-\ram operator translation rules.
    We assume register names are created from fresh symbols and never conflict.
    We use $\cdot$ to denote sequential composition of instructions and $\gets$ to denote assignment.
    Translation proceeds in the context of a database that is partitioned into three components: $F_T^\text{stable}$, $F_T^\text{recent}$, and $F_T^\Delta$.
    This enables semi-naive evaluation, as discussed in Section~\ref{sec:apm-evaluation}.
    }
    \label{fig:ram-apm-translation}
\end{figure*}

\section{\apm Evaluation}

\SetKwComment{Comment}{/* }{ */}

\RestyleAlgo{ruled}
\begin{algorithm}[t]
\footnotesize
\caption{How to execute a \ram program via compilation to and execution of \apm.}
\label{alg:apm-execution}
\KwData{Database $F_T$, RAM program $\overline{\phi}$.}
\KwResult{
$F_T$ updated to reflect the result of evaluating $\overline{\phi}$.
}
\For{$\phi$ in $\overline{\phi}$}{
  ${\it instructions} \gets {\tt compile}(\phi)$\;
  $F_T^\text{stable}$, $F_T^\text{recent}$, $F_T^{\Delta}$ $\gets$ $\emptyset$, $F_T$, $\emptyset$\;
  ${\it size} \gets |F_T^\text{stable}|$\;
  \While{true}{
    \For{$i$ in {\it instructions}}{
      execute $i$
    }
    ${\it size}_{\textnormal{new}} \gets |F_T|$\;
    \If{${\it size}_{\textnormal{new}} = {\it size}$}{
      \textbf{break}\;
    }
    ${\it size} \gets {\it size}_{\textnormal{new}}$\;
  }
  $F_T \gets F_T^\text{stable}$
}

\end{algorithm}

Details on executing a \ram program via compilation to \apm are provided in Algorithm \ref{alg:apm-execution}.

\end{document}